\documentstyle[prb,floats,aps,epsf]{revtex}
\begin{document}
\twocolumn
\title{Temperature dependence of the resistivity \\ in the double-exchange model}
\author{Satoshi Ishizaka}
\address{NEC Fundamental Research Laboratories, \\
34 Miyukigaoka, Tsukuba, Ibaraki, 305, Japan}
\author{Sumio Ishihara}
\address{Institute for Materials Research, Tohoku University, \\ Sendai,
980-8577, Japan}
\date{\today}
\maketitle
%%%%%%%%%%%%%%%%%%%%%%%%%%%%%%%%%%%%%%%%%%%%%%%%%%%%%%%%%%%%%%%%%%%%%%%%%%%%%%%
\begin{abstract}
The resistivity around the ferromagnetic transition temperature
in the double exchange model is studied by the Schwinger boson approach.
The spatial spin correlation responsible for scattering of conduction
electrons are taken into account by adopting the memory function formalism. 
Although the correlation shows a peak lower than the transition temperature,
the resistivity in the ferromagnetic state monotonically increases with
increasing temperature due to a variation of the electronic state of the
conduction electron.
In the paramagnetic state, the 
resistivity is dominated by the short range correlation of scattering 
and is almost independent of the temperature. 
It is attributed to a cancellation between the
nearest-neighbor spin correlation, the fermion bandwidth,
and the fermion kinetic energy. 
This result implies the importance of the temperature dependence of the 
electronic states of the conduction electron 
as well as the localized spin states in both ferromagnetic and paramagnetic
phases. 
\end{abstract}
%%%%%%%%%%%%%%%%%%%%%%%%%%%%%%%%%%%%%%%%%%%%%%%%%%%%%%%%%%%%%%%%%%%%%%%%%%%%%%%
\narrowtext
%%%%%%%%%%%%%%%%%%%%%%%%%%%%%%%%%%%%%%%%%%%%%%%%%%%%%%%%%%%%%%%%%%%%%%%%%%%%%%%
The recent discovery of colossal magnetoresistance (CMR)
\cite{chah,hel,toku,jin} 
has revived interest in perovskite manganites such as La$_{1-x}$Sr$_x$MnO$_3$.
It is widely accepted that significant changes in the transport properties, 
as well as CMR, are observed around the transition between the ferromagnetic
phase and the paramagnetic one. 
More than 40 years ago, 
Zener \cite{Zener51a} proposed a double-exchange (DE) interaction to explain
the correlation between electrical conduction and the ferromagnetism, in which
the spin of a conduction electron and a localized core spin ($\vec S$)
on the same site are strongly coupled by Hund's rule.
Since the hopping amplitude of the electron to the
neighboring sites is maximum when the two neighboring core spins
are parallel, the ferromagnetic metallic state is achieved by gaining
the kinetic energy of the conduction electron.
\cite{Zener51a,Anderson55a,deGennes60a}
These concepts were settled as the so-called DE model and 
the magnetic and transport properties in this model 
have been investigated intensively and extensively.   
\par
%%%%%%%%%%%%%%%%%%%%%%%%%%%%%%%%%%%%%%%%%%%%%%%%%%%%%%%%%%%%%%%%%%%%%%%%%%%%%%%
%
One of the main interest in this research field is the temperature 
dependence of the electrical resistivity. 
The resistivity in the DE model was studied in a mean-field theory
by Kubo and Ohata, \cite{KuboOhata72a}
and similar results have been reproduced by a dynamical mean-field theory by
Furukawa. \cite{Furukawa94a}
In these calculations, however, the spatial correlation of the core
spins is not included properly,  
although it is pointed out that 
it plays a crucial role in the electric transport 
near the Curie temperature ($T_{c}$). \cite{deGennes58a,Fisher68a}
The short range spin correlation was only considered 
in Ref. \onlinecite{KuboOhata72a},   
and the spatial correlation was neglected in Ref. \onlinecite{Furukawa94a}, 
where the dynamical fluctuation was taken into account. 
Millis {\it et al.} discussed a possibility
that the behavior of the resistivity is greatly modified when the 
spatial correlation of the core spins 
is properly taken into account. \cite{Millis95a}
They showed that the resistivity is still increased below $T_{c}$
with decreasing the temperature. 
Being based on the calculated results which disagree with the experimental
one, they concluded that the additional gradients, such as, the Jahn-Teller 
effect, are necessary to reproduce the observed behaviors. 
\par
%%%%%%%%%%%%%%%%%%%%%%%%%%%%%%%%%%%%%%%%%%%%%%%%%%%%%%%%%%%%%%%%%%%%%%%%%%%%%%%
In this paper, we calculate the temperature dependence of the 
resistivity in the DE model by the Schwinger boson approach.
In order to include 
the effects of the spatial correlation of the core spins properly, 
we adopt the memory function formalism which 
was also used in Ref.\ \onlinecite{Millis95a}. 
In addition, the temperature dependence of the electronic structure 
is determined self-consistently together with that of the core spins. 
The calculated resistivity monotonically decreases with decreasing
temperature  in the ferromagnetic states 
and does not show a peak below $T_{c}$,  
although the spin correlation has its maximum at $T<T_{c}$. 
\par
%%%%%%%%%%%%%%%%%%%%%%%%%%%%%%%%%%%%%%%%%%%%%%%%%%%%%%%%%%%%%%%%%%%%%%%%%%%%%%%
The Hamiltonian of DE model in the limit of strong Hund's coupling
is given by the Schwinger boson representation as follows,  
\begin{equation}
{\cal H}=-\frac{t}{2S_R}\sum_{\langle ij \rangle \sigma}
\Big[ b_{i\sigma}^\dagger b_{j\sigma} f_i^\dagger f_j + \hbox{h.c.} \Big]
\end{equation}
with the local constraint
$\sum_\sigma b_{i\sigma}^\dagger b_{i\sigma} - f_i^\dagger f_i\!=\!2S$
at every lattice site $i$. 
Here, $b_{i\sigma}$ ($\sigma=\uparrow,\downarrow$) is a boson
and $f_i$ is a spinless fermion operator,
$S_R\!=\!S+(1-x)/2$, \cite{Arovas88a,Sarker96a,Arovas98a}
and $x$ is the doping concentration of holes
($\langle f_i^\dagger f_i \rangle\!=\!1-x$).
We shall exclusively consider the case of $S\!=\!\frac{3}{2}$. 
A transition at $T_c$ from a ferromagnetic state to a paramagnetic state
(described as Bose condensation of Schwinger bosons)
was investigated by using a mean field Hamiltonian
%as follows,
\cite{Sarker96a} 
\begin{eqnarray}
{\cal H}_{MF}&=&-\frac{Bt}{S_R}\sum_{\langle ij \rangle}
\Big[ f_i^\dagger f_j + \hbox{h.c.} \Big] \cr
&&-\frac{Dt}{2S_R}\sum_{\langle ij \rangle \sigma}
\Big[ b_{i\sigma}^\dagger b_{j\sigma} + \hbox{h.c.} \Big]
\label{eq: Mean Field Hamiltonian}
\end{eqnarray}
with a global constraint
\begin{equation}
\langle \sum_\sigma b_{i\sigma}^\dagger b_{i\sigma}\rangle\!=\!2S_R,
\label{eq: Global Constraint}
\end{equation}
where $B$ and $D$ are given by
$\frac{1}{2}\sum_\sigma\langle b_{i\sigma}^\dagger b_{j\sigma} \rangle$
and $\langle f_i^\dagger f_j \rangle$, respectively, 
and both are determined
self-consistently.
This mean field treatment, however, leads to an additional transition 
at slightly higher temperature than $T_c$ (about $1.4 T_c$) into an
artifact state in which $B\!=\!D\!=\!0$. \cite{Arovas98a}
As a result, above $T_c$ the fermion bandwidth ($W_f\equiv12Bt/S_R$) rapidly
decreases. This decrease obviously causes a misleading diverging increase of
the resistivity.
\par
%%%%%%%%%%%%%%%%%%%%%%%%%%%%%%%%%%%%%%%%%%%%%%%%%%%%%%%%%%%%%%%%%%%%%%%%%%%%%%%
In the present study, we assume that 
Eq.\ (\ref{eq: Mean Field Hamiltonian}) itself has a suitable form as a
mean field Hamiltonian, but in order to avoid the difficulty mentioned above,
the fermion bandwidth $B$ is determined as
\begin{eqnarray}
B&\equiv&\frac{1}{2}
\Big\langle\sqrt{|\sum_\sigma b_{i\sigma}^\dagger b_{j\sigma}|^2}\Big\rangle\cr
&=&\frac{S_R}{\sqrt{2}}
\Big\langle \sqrt{1+(\frac{1}{2S_R^2}\sum_{\sigma\sigma'}
b_{i\sigma}^\dagger b_{i\sigma'} b_{j\sigma'}^\dagger b_{j\sigma}-1)
+\frac{1}{S_R}} \Big\rangle \cr
&\approx&\frac{S_R}{\sqrt{2}}\Big[\frac{1}{2}+\frac{1}{4S_R^2}
\sum_{\sigma\sigma'}
\langle
b_{i\sigma}^\dagger b_{i\sigma'}b_{j\sigma'}^\dagger b_{j\sigma}
\rangle
\Big] 
\label{eq: Determination of B}
\end{eqnarray}
together with
\begin{equation}
D\!\equiv\!\langle f_i^\dagger f_j \rangle
\end{equation}
in a self-consistent manner.
Here, we ignore the Berry's phase in
the electron hopping and use the fact that
$\sum_{\sigma\sigma'}b_{i\sigma}^\dagger b_{i\sigma'} b_{j\sigma'}^\dagger b_{j\sigma}\rightarrow2S_R^2$ in the high-temperatures limit.
The approximation in Eq.\ (\ref{eq: Determination of B}) corresponds to the expansion with respect to $\vec S_i \cdot \vec S_j$.
It should be noted that the fermion bandwidth 
obtained in Eq. (4) remains finite
for $T\!\rightarrow\!\infty$ as expected, while 
$\frac{1}{2}\sum_\sigma\langle b_{i\sigma}^\dagger b_{j\sigma} \rangle$
characterizing a nearest-neighbor magnetic correlation (denoted by $C$)
vanishes for $T\!\rightarrow\!\infty$ in this model.
Therfore, the behavior obtained in the above formulas is physically reasonable.
\par
%%%%%%%%%%%%%%%%%%%%%%%%%%%%%%%%%%%%%%%%%%%%%%%%%%%%%%%%%%%%%%%%%%%%%%%%%%%%%%%
\begin{figure}
\epsfxsize=7.0cm
\centerline{\epsfbox{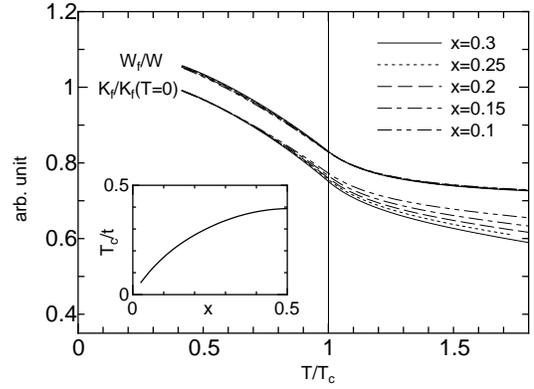}}
\caption{
The fermion bandwidth normalized by bare bandwidth ($W_f/W$) and
the fermion kinetic energy normalized by the zero-temperature value
($K_f/K_f(T\!=\!0)$)
as a function of $T/T_c$ for several doping concentrations.
The inset shows the transition temperature from a ferromagnetic state
to a paramagnetic state as a function of the doping concentration.
}
\label{fig: Band Width}
\end{figure}
%%%%%%%%%%%%%%%%%%%%%%%%%%%%%%%%%%%%%%%%%%%%%%%%%%%%%%%%%%%%%%%%%%%%%%%%%%%%%%%
The Curie temperature ($T_c$) in this model
as a function of the doping concentration is shown in the inset of
Fig.\ \ref{fig: Band Width}.
Since $T_c$ is determined by Eq.\ (\ref{eq: Global Constraint}),
the results are essentially independent on the formula of $B$ and are
the same as those in Ref.\ \onlinecite{Sarker96a}.
The transition temperature is scaled independent on $x$ by $D$ as
$T_c/t\!\approx\!2.5D$. 
\par
%%%%%%%%%%%%%%%%%%%%%%%%%%%%%%%%%%%%%%%%%%%%%%%%%%%%%%%%%%%%%%%%%%%%%%%%%%%%%%%
The temperature dependencies of the fermion bandwidth normalized by a bare
band-width ($W\!=\!12t$) are plotted in Fig.\ \ref{fig: Band Width}.
At the transition temperature, the bandwidth does not directly depend on the
doping
concentration (it depends on $x$ only through $S_R$). 
This is because both the chemical
potential and
the condensate density of bosons are zero, and $Dt/T_c$, 
which is independent of $x$, is
only a parameter in Eq.\ (\ref{eq: Determination of B}). \cite{Sarker89a}
Further, $W_f/W$ approaches $1/\sqrt{2}$ in an infinite temperature
limit.
As a result, the behavior of the fermion bandwidth as a function of
$T/T_c$ is almost universally independent of $x$.
On the other hand, the behavior of the kinetic energy of the
fermion ($K_f\!\equiv\! Bt\langle f_i^\dagger f_j \rangle/S_R$)
as a function of $T/T_c$ (Fig.\ \ref{fig: Band Width}) changes
depending on $x$.
%%%%%%%%%%%%%%%%%%%%%%%%%%%%%%%%%%%%%%%%%%%%%%%%%%%%%%%%%%%%%%%%%%%%%%%%%%%%%%%
It is important that the bandwidth varies even in the disordered-spin regime
of $T\!>\!T_c$.
This behavior agrees well with the result in Ref.\ \onlinecite{Calderon98a}.
In fact, the bandwidth at $T_c$ is 1.16 times bigger than
that in the $T\!\rightarrow\!\infty$ limit.
\par
%%%%%%%%%%%%%%%%%%%%%%%%%%%%%%%%%%%%%%%%%%%%%%%%%%%%%%%%%%%%%%%%%%%%%%%%%%%%%%%
The resistivity as a function
of the temperature is calculated by the memory function method,
\cite{Mori65a,Getze72a}
where the lowest order fluctuation from the mean field can be
included automatically.
In this lowest-order perturbational treatment, 
a static approximation for Schwinger bosons
is appropriate.
This is because the bandwidth of Schwinger bosons is much smaller than
that of the fermion ($D\!\ll\!B$) and the effects of the quantum fluctuation
of bosons can be negligible for
$T\lower -0.3ex \hbox{$>$} \kern -0.75em \lower0.7ex\hbox{$\sim$}0.5T_c$ where
$Dt/(2S_RT)\!\ll\!1$.
The memory function is evaluated to leading order in $1/S_R$,
and thus, the resistivity is written as
\begin{equation}
\rho=\frac{\hbar^2a}{2 e^2 K_f \tau} , 
\label{eq:Resistivity0}
\end{equation}
with
\begin{eqnarray}
\frac{1}{\tau}&=&\frac{\pi t^4}{8\hbar K_fS_R^4}\sum_{i}\Gamma(\vec R_i)
\frac{1}{N^2}\sum_{\vec p_1 \vec p_2}
\Bigg(
-\frac{\partial f(\varepsilon_{\vec p_1})}{\partial \varepsilon_{\vec p_1}}
\Bigg) \cr
&\times&
\delta(\varepsilon_{\vec p_1}-\varepsilon_{\vec p_2})
(e^{i\kappa_x}-1)(e^{-i\kappa_x}-1) e^{-i\vec \kappa \cdot \vec R_i},
\label{eq: Resistivity}
\end{eqnarray}
where $\Gamma(\vec R_i)$ represents the spatial spin correlation 
defined as 
\begin{equation}
\Gamma(\vec R_i)\equiv\sum_{\sigma \sigma' \rho \rho'}
\langle b_{0\sigma}^\dagger b_{x\sigma} b_{x\sigma'}^\dagger b_{0\sigma'}
b_{i\rho}^\dagger b_{i+x\rho} b_{i+x\rho'}^\dagger b_{i\rho'} \rangle. 
\label{eq: Gamma}
\end{equation}
Here, 
$f(\varepsilon_{\vec p})$ is the Fermi distribution function of the
spinless fermion,
$\vec\kappa\!=\!\vec p_1-\vec p_2$ is the momentum transfer of the fermion
due to scattering, and the average $\langle\ldots\rangle$ is evaluated by the mean field
Hamiltonian in Eq.\ (\ref{eq: Mean Field Hamiltonian}).
The same expression given by spin-variables has been derived
in Ref.\ \onlinecite{Millis95a}.
It should be noted that in this model the actual value of the resistivity
in units of $ha/e^2$ does not depend on that of $t$.
This is because $\varepsilon_{\vec p}$ and $K_f$ are scaled by $t$, and 
$\tau$ is scaled by $1/t$; thus, $K_f\tau$ becomes independent of $t$.
\par
%%%%%%%%%%%%%%%%%%%%%%%%%%%%%%%%%%%%%%%%%%%%%%%%%%%%%%%%%%%%%%%%%%%%%%%%%%%%%%%
\begin{figure}
\epsfxsize=7.0cm
\centerline{\epsfbox{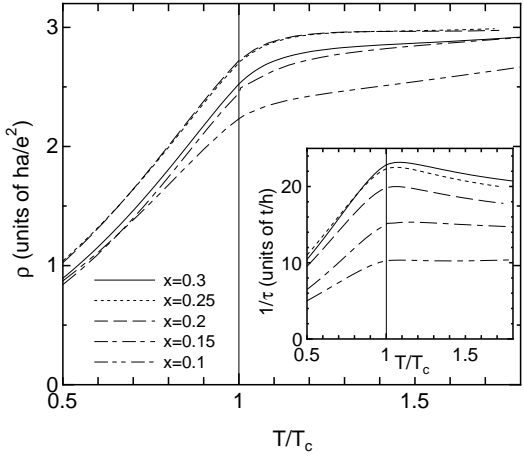}}
\caption{
Calculated resistivity as a function of $T/T_c$ for several doping
concentrations.
The results for $x\!=\!0.2$ and $x\!=\!0.25$ are almost the same and are
overlapping.
The corresponding inverse relaxation time $(1/\tau)$ is shown in the inset.
}
\label{fig: Resistivity}
\end{figure}
%%%%%%%%%%%%%%%%%%%%%%%%%%%%%%%%%%%%%%%%%%%%%%%%%%%%%%%%%%%%%%%%%%%%%%%%%%%%%%%
Results of the calculated resistivity ($\rho$) are shown in
Fig.\ \ref{fig: Resistivity} as a function of the temperature ($T$) for several
doping concentrations.
The resistivity monotonically increases with increasing $T$ in
a ferromagnetic state for all doping concentrations.
In a paramagnetic state, the resistivity still weakly increases,
i.e., metallic for $x\!<\!0.2$.
In the case of $x\!=\!0.2$ to 0.25, the $T$-dependence almost
vanishes
for $T\lower -0.3ex \hbox{$>$} \kern -0.75em \lower0.7ex\hbox{$\sim$}1.2T_c$.
For a further high doping concentration, $x\!>\!0.2$, the resistivity comes to
weakly increase again.
On the other hand, the corresponding inverse relaxation time ($1/\tau$) 
shown in the inset of the figure shows a different $T$-dependence:
it decreases in the paramagnetic state for all cases of $x$.
This difference clearly indicates the importance of the $T$-dependence
of $K_f$ since $\rho\propto 1/(K_f\tau)$.
\par
%%%%%%%%%%%%%%%%%%%%%%%%%%%%%%%%%%%%%%%%%%%%%%%%%%%%%%%%%%%%%%%%%%%%%%%%%%%%%%%
\begin{figure}
\epsfxsize=7.0cm
\centerline{\epsfbox{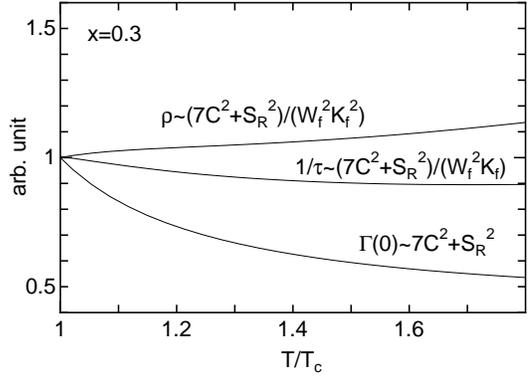}}
\caption{
An approximate temperature dependence of
$\Gamma(0)$, $1/\tau$, and $\rho$ obtained by Eqs. 
(10)-(12), respectively, 
in the paramagnetic state for $x\!=\!0.3$.
Each result is normalized by the value at $T_c$.
}
\label{fig: T-Dependence}
\end{figure}
%%%%%%%%%%%%%%%%%%%%%%%%%%%%%%%%%%%%%%%%%%%%%%%%%%%%%%%%%%%%%%%%%%%%%%%%%%%%%%%
These behaviors in the paramagnetic state can be understood in
terms of Fisher and Langer's scheme, \cite{Fisher68a}
although it was originally proposed to analyze the electron transport in 
the transition metal ferromagnets.
The inverse relaxation time in the DE model can be rewritten as
\cite{Fisher68a}
\begin{equation}
\frac{1}{\tau} \propto \frac{D(E_F)^2}{K_f} \sum_{i} \Gamma(\vec R_i)
f(\vec R_i),
\end{equation}
where $f(\vec R_i)$ is the
decaying oscillatory function,
and $D(E_F)$ is the density of states at the Fermi level.
In the case of the transition metal ferromagnets, $\Gamma(0)$ does not
depend on $T$ in the paramagnetic state; thus, the temperature dependent
part of the resistivity is wholly determined by $\Gamma(\vec R_i)$ for 
$\vec R_i\ne0$. \cite{Fisher68a}
In the case of the DE model, however, the term with 
$\vec R_i\!=\!0$ depends on $T$ through $C$ as
\begin{equation}
\Gamma(0)\sim8S_R^2(7C^2+S_R^2) \hbox{~for~} S_R\gg1 \ , 
\label{eq:gamma0}
\end{equation}
where $C$ is the correlation function between nearest neighboring spins. 
It gives a dominant contribution in $1/\tau$. 
Further, $D(E_F)$ ($\propto\!1/W_f$) and $K_f$ also depend on $T$;
therefore, the temperature dependence of $1/\tau$ and $\rho$ are given by 
\begin{equation}
\frac{1}{\tau} \propto \frac{7C^2+S_R^2}{W_f^2 K_f}, 
\label{eq: Cancelation0}
\end{equation}
%
%and 
%
\begin{equation}
\rho\propto\frac{7C^2+S_R^2}{W_f^2 K_f^2}.
\label{eq: Cancelation}
\end{equation} 
%respectively. 
These quantities for $x\!=\!0.3$ are plotted in Fig.\ \ref{fig: T-Dependence}
and roughly agree with the calculated results shown in
Fig.\ \ref{fig: Resistivity}.
The rounding of the calculated $\rho$ at very close to $T_c$, however, must
come from $\Gamma(\vec R_i)$ at $\vec R_i\!\ne\!0$.
\par
\begin{figure}
\epsfxsize=7.0cm
\centerline{\epsfbox{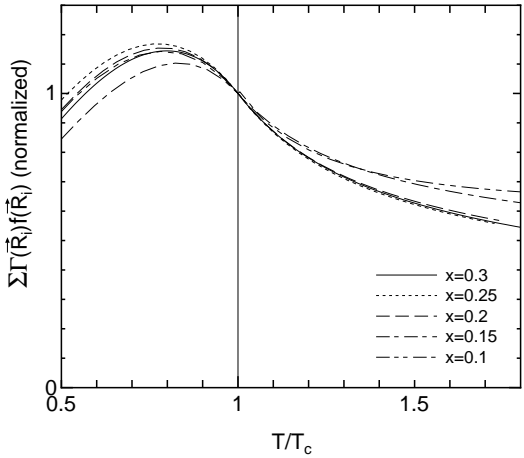}}
\caption{
The temperature dependence of $\sum_i\Gamma(\vec R_i)f(\vec R_i)$
for several doping concentrations.
Results are normalized by the value at $T_c$.
}
\label{fig: Memory Function}
\end{figure}

%%%%%%%%%%%%%%%%%%%%%%%%%%%%%%%%%%%%%%%%%%%%%%%%%%%%%%%%%%%%%%%%%%%%%%%%%%%%%%%
It is worth to note that 
the calculated results in the temperature dependence of the resistivity 
are different from those in
Ref.\ \onlinecite{Millis95a}, 
although the memory function formalism is adopted in both cases.
The discrepancy is attributed to the fact that 
in the present calculation 
the temperature dependence of the electronic structure, that is, 
$W_f$ and $K_f$ (shown in Fig.\ \ref{fig: Band Width}), 
are taken into account, as well as that of  
$\sum_i\Gamma(\vec R_i)f(\vec R_i)$. 
In fact, $\sum_i\Gamma(\vec R_i)f(\vec R_i)$ plotted in 
Fig.\ \ref{fig: Memory Function}
shows a peak at $T<T_c$ which is smeared out in the resistivity
due to the variation of the electronic structure.
The peak structure originates from the process in which bosons
in the condensate part are scattered to the non-condensate part, and vice
versa. \cite{Millis95a}
It should be noted that,
since the number of condensate bosons is macroscopically large,
the lowest order perturbational treatment for such scattering processes
might overestimate the scattering amplitude, and the resistivity
might become slightly smaller than the present results because of higher order
perturbations.
In any case, the resistivity is unlikely to show the peak in the ferromagnetic
state.
\par
%%%%%%%%%%%%%%%%%%%%%%%%%%%%%%%%%%%%%%%%%%%%%%%%%%%%%%%%%%%%%%%%%%%%%%%%%%%%%%%
%%%%%%%%%%%%%%%%%%%%%%%%%%%%%%%%%%%%%%%%%%%%%%%%%%%%%%%%%%%%%%%%%%%%%%%%%%%%%%%
On the other hand, our results are similar to the previous results
by Kubo and Ohata \cite{KuboOhata72a}
except for the discontinuity in $d\rho/dT$ at $T_c$.
In Ref.\ \onlinecite{KuboOhata72a},
only the shortest correlation of scattering ($\vec R_i\!=\!0$) is included
and this is a suitable approximation as discussed above.
It should be noted, however, that the physical mechanism leading to the
weak temperature dependence of the resistivity in the paramagnetic state is
completely different.
The mechanism is a result of the cancellation between the band-width ($W_f$), 
the kinetic energy ($K_f$), and the nearest-neighbor spin correlation
($C$) in a somewhat complicated manner (Eq.\ (\ref{eq: Cancelation}))
in our calculations.
On the other hand, these quantities used in 
Ref.\ \onlinecite{KuboOhata72a} are temperature independent in the paramagnetic
state.
Further, the discontinuity in $d\rho/dT$ at $T_c$ comes from the sudden
change of $W_f$,
which is absent in our results since 
the contribution from $\Gamma(\vec R_i)$ with $R_i\ne 0$ is not neglected near $T_c$ and 
$W_f$ is a smooth function of $T$.
Although our results are also similar to the results by Furukawa,
\cite{Furukawa94a}
it is likely to be just a coincidence because the spatial correlation of core
spins between different sites are neglected and 
scattering processes responsible for the resistivity are different.
%%%%%%%%%%%%%%%%%%%%%%%%%%%%%%%%%%%%%%%%%%%%%%%%%%%%%%%%%%%%%%%%%%%%%%%%%%%%%%%
It should be noted further that our results qualitatively agree with the
recent results by Monte Carlo simulations. \cite{Calderon98b}
For $a\!=4\!\AA$, at $T_c$, we obtain $\rho\!\sim\!2\times10^{-3}$ $\Omega$cm,
whose order also agrees with these results.
\par
%%%%%%%%%%%%%%%%%%%%%%%%%%%%%%%%%%%%%%%%%%%%%%%%%%%%%%%%%%%%%%%%%%%%%%%%%%%%%%%
To conclude, using the Schwinger boson approach, we have calculated the
resistivity in the double-exchange model.
In this approach, 
the fermion bandwidth has been determined by the absolute value of the hopping
amplitude giving a physically reasonable temperature dependence in contrast to
the conventional Schwinger boson approach.
The resistivity monotonically increases with increasing temperature
in the ferromagnetic state, which is different from the previous results in
Ref.\ \onlinecite{Millis95a}.
In the paramagnetic state,
the resistivity is dominated by the short range correlation of scattering.
Although the behavior slightly changes depending on the doping concentration,
the temperature dependence almost vanishes due to
a cancellation between the nearest-neighbor magnetic correlation,
the fermion bandwidth, and the fermion kinetic energy.
These results imply the importance of the temperature dependence of 
the electronic structure of the conduction electron in 
the both ferromagnetic and paramagnetic phases.
\par
%%%%%%%%%%%%%%%%%%%%%%%%%%%%%%%%%%%%%%%%%%%%%%%%%%%%%%%%%%%%%%%%%%%%%%%%%%%%%%%
The results agree with the experiments for ``higher'' doping concentrations
$x\!\sim\!0.3$, where experimentally observed resistivity is relatively low
as a whole ($\rho\!\sim\!5\times10^{-3}$ $\Omega$cm at $T_c$) and shows a
metallic behavior in the paramagnetic state.
For lower concentrations $x\!\sim\!0.2$, however, the experimentally observed
singular behavior around the transition temperature and the insulating
behavior in the paramagnetic state are not reproduced in our calculations.
Other effects might play a crucial role in cooperation with the double
exchange mechanism for the system with such the lower concentration.
\par
%%%%%%%%%%%%%%%%%%%%%%%%%%%%%%%%%%%%%%%%%%%%%%%%%%%%%%%%%%%%%%%%%%%%%%%%%%%%%%%
One of the authors (Ishizaka) would like to thank T. Hiroshima for helpful
discussions.
%%%%%%%%%%%%%%%%%%%%%%%%%%%%%%%%%%%%%%%%%%%%%%%%%%%%%%%%%%%%%%%%%%%%%%%%%%%%%%%
%
%%%%%%%%%%%%%%%%%%%%%%%%%%%%%%%%%%%%%%%%%%%%%%%%%%%%%%%%%%%%%%%%%%%%%%%%%%%%%%%

%%%%%%%%%%%%%%%%%%%%%%%%%%%%%%%%%%%%%%%%%%%%%%%%%%%%%%%%%%%%%%%%%%%%%%%%%%%%%%%
%%
%%%%%%%%%%%%%%%%%%%%%%%%%%%%%%%%%%%%%%%%%%%%%%%%%%%%%%%%%%%%%%%%%%%%%%%%%%%%%%%

\vspace{-0.5cm}
\begin{thebibliography}{20}
\vspace{-1.5cm}
\bibitem{chah}
K. Chahara, T. Ohono, M. Kasai, Y. Kanke, and Y. Kozono, Appl. Phys. Lett. 
{\bf 62}, 780  (1993). 

\bibitem{hel}
R. von Helmolt, J. Wecker, B. Holzapfel, L. Schultz, and K. Samwer,
Phys. Rev. Lett. {\bf 71}, 2331 (1993). 

\bibitem{toku}
Y. Tokura, A. Urushibara, Y. Moritomo, T. Arima, A. Asamitsu, G. Kido,
and N. Furukawa, J. Phys. Soc. Jpn. {\bf 63}, 3931 (1994).

\bibitem{jin}
S. Jin, T. H. Tiefel, M. McCormack, R. A. Fastnacht, R. Ramesh,
and L. H. Chen, Science, {\bf 264}, 413 (1994)

\bibitem{Zener51a}
C. Zener, Phys. Rev. {\bf 82},  403  (1951).

\bibitem{Anderson55a}
P.~W. Anderson and H. Hasegawa, Phys. Rev. {\bf 100},  675  (1955).

\bibitem{deGennes60a}
P.~G. de~Gennes, Phys. Rev. {\bf 118},  141  (1960).

\bibitem{KuboOhata72a}
K. Kubo and N. Ohata, J. Phys. Soc. Jpn. {\bf 33},  21  (1972).

\bibitem{Furukawa94a}
N. Furukawa, J. Phys. Soc. Jpn. {\bf 63},  3214  (1994).

\bibitem{deGennes58a}
P.~G. de~Gennes and J. Friedel, J. Phys. Chem. Solids {\bf 4},  71  (1958).

\bibitem{Fisher68a}
M.~E. Fisher and J.~S. Langer, Phys. Rev. Lett. {\bf 20},  665  (1968).

\bibitem{Millis95a}
A.~J. Millis, P.~B. Littlewood, and B.~I. Shraiman, Phys. Rev. Lett. {\bf 74},
  5144  (1995).

\bibitem{Arovas88a}
D.~P. Arovas and A. Auerbach, Phys. Rev. B {\bf 38},  316  (1988).

\bibitem{Sarker96a}
S.~K. Sarker, J. Phys.: Condens. Matter {\bf 8},  L515  (1996).

\bibitem{Arovas98a}
D.~P. Arovas and F. Guinea, Phys. Rev. B {\bf 58},  9150  (1998).

\bibitem{Calderon98a}
M.~J. Calder{\"o}n and L. Brey, Phys. Rev. B {\bf 58},  3286  (1998).

\bibitem{Sarker89a}
S. Sarker, C. Jayaprakash, H.~R. Krishnamurthy, and M. Ma, Phys. Rev. B
{\bf 40},  5028  (1989).

\bibitem{Mori65a}
H. Mori, Prog. Theor. Phys. {\bf 34},  399  (1965).

\bibitem{Getze72a}
W. G{\"o}tze and P. W{\"o}lfle, Phys. Rev. B {\bf 6},  1226  (1972).

\bibitem{Calderon98b}
M.~J. Calder{\"o}n, J.~A. Verg{\'e}s, and L. Brey, cond-mat/9806157.
\end{thebibliography}
\end{document}